\documentclass[a4paper]{amsart}
\title{A note on Integer Factorization Using Lattices}
\author[Antonio Vera]{Antonio Vera\\ \\
CNRS/INRIA/Nancy-Universit\'{e}}
\usepackage{abbrev}
\usepackage{stmaryrd}

\newtheorem{lemma}{Lemma}
\newtheorem{theorem}{Theorem}

\theoremstyle{remark}
\newtheorem{remk}{Remark}
\usepackage[pdftex]{hyperref}
\begin{document}
\maketitle
\begin{abstract}
  We revisit Schnorr's lattice-based integer factorization algorithm, now with an effective point of view. We present effective versions of Theorem 2 of \cite{Sch93}, as well as new properties of the Prime Number Lattice bases of Schnorr and Adleman.
\end{abstract}
\tableofcontents
\section{Introduction}
\label{sec:intro}
\begin{figure}
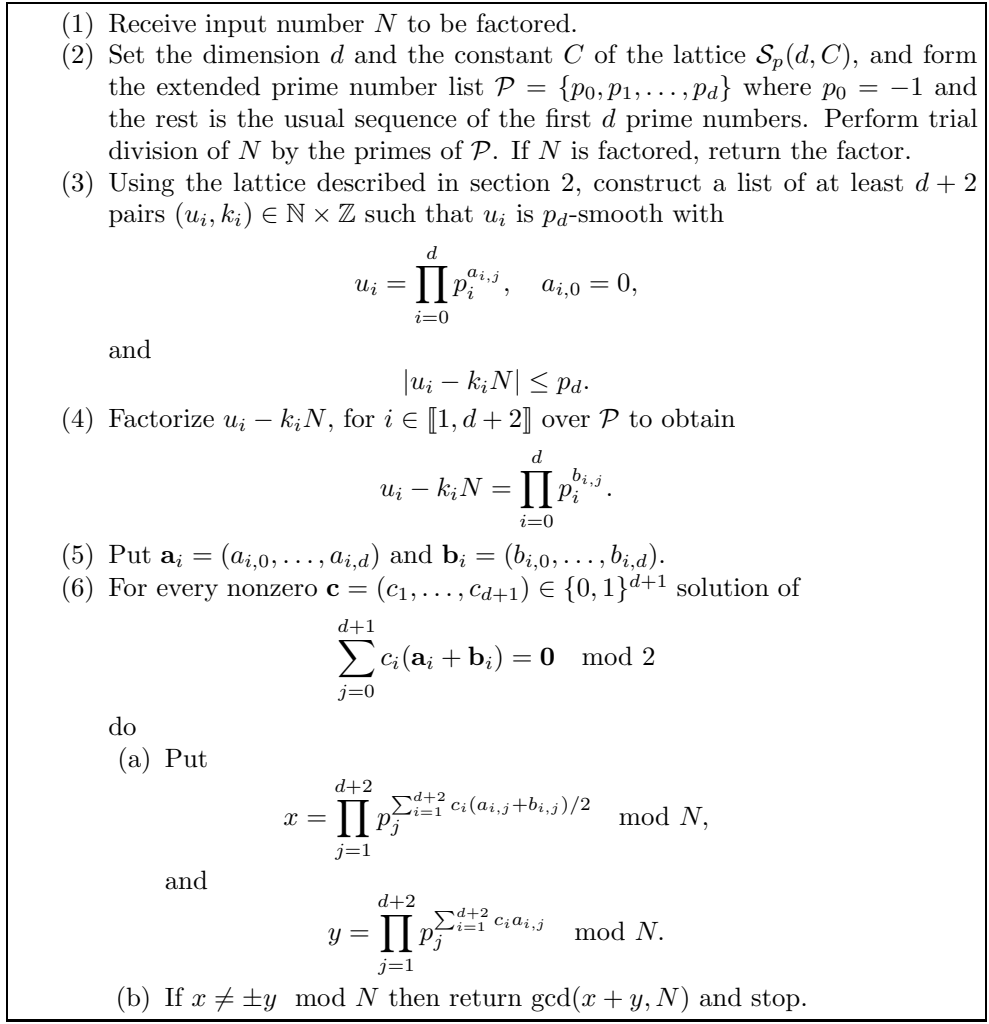

  \centering
  \framebox{\begin{minipage}{1.0\linewidth}
      \begin{enumerate}
      \item Receive input number $N$ to be factored.
      \item Set the dimension $d$ and the constant $C$ of the lattice $\cS_p(d,C)$, and form the extended prime number list
        $\cP=\{p_0,p_1,\dots,p_d\}$ where $p_{0}=-1$ and the rest is the usual sequence of the first $d$ prime numbers. Perform trial division of $N$ by the primes of $\cP$. If $N$ is factored, return the factor.
      \item \label{it:relsearch} Using the lattice described in section
        \ref{sec:prime-number-lattice}, construct a list of at least
        $d+2$ pairs $(u_i,k_i)\in \bN\times \bZ$ such that $u_i$ is $p_d$-smooth
        with
        $$ u_i =  \prod_{i=0}^d p_i^{a_{i,j}},\quad a_{i,0}=0, $$
        and
        $$ |u_i-k_iN|\leq p_d. $$
      \item Factorize $u_i-k_iN$, for $i\in \llbracket 1,d+2 \rrbracket$
        over $\cP$ to obtain
        $$ u_i-k_iN = \prod_{i=0}^d p_i^{b_{i,j}}.  $$
      \item Put $\na_i=(a_{i,0},\dots,a_{i,d})$ and $\nnb_i =
        (b_{i,0},\dots, b_{i,d})$.
      \item \label{it:linalg}For every nonzero $\nc=(c_1,\dots,c_{d+1})
        \in \{0,1\}^{d+1}$ solution of
        $$ \sum_{j=0}^{d+1} c_i(\na_i+\nnb_i) = \mathbf{0}\mod 2 $$
        do
        \begin{enumerate}
        \item Put
          $$ x = \prod_{j=1}^{d+2} p_j^{\sum_{i=1}^{d+2} c_i(a_{i,j}+b_{i,j})/2}\mod N, $$
          and
          $$ y = \prod_{j=1}^{d+2} p_j^{\sum_{i=1}^{d+2} c_i a_{i,j}}\mod N. $$ % \left( = \prod_{j=1}^{d+2} p_j^{\sum_{i=1}^{d+2} c_i b_{i,j}}\mod N \right)
        \item If $x\neq \pm y\mod N$ then return $\text{gcd}(x+y,N)$ and stop.
        \end{enumerate}      
      \end{enumerate}
    \end{minipage}}
  \caption{Outline of Schnorr's algorithm}
  \label{fig:schalg}
\end{figure}

Let $N\geq 1$ be a composite integer that we want to factor. The \emph{congruence of squares method} consists of finding $x,y\in \bZ$ such that
\begin{equation}
 x^2\equiv y^2 \mod N\label{eq:congsq}
\end{equation}
with $x\not\equiv\pm y \mod N$, and factor $N$ by computing $\text{gcd}(x+y,N)$. Although this is a heuristic method, it works pretty well in practice and one can show under reasonable hypotheses (see \cite[page 268, remark (5)]{PRIMES}) that for random $x,y$ satisfying (\ref{eq:congsq}), one has $x\not\equiv\pm y \mod N$ with probability $\geq 1/2$. This report considers an algorithm based on this philosophy, namely Schnorr's algorithm \cite{Sch93}, whose outline is given in figure \ref{fig:schalg}.

Call $B$-smooth an integer free of prime factors $>B$, and let $p_i$ be the $i$-th prime number. Fix some $d\geq 1$ and suppose that $N$ is free of prime factors $\leq p_d$.  The core computational task of the algorithm consists in finding $d+2$ integer quartets $(u,v,k,\gamma)$, with $u, v$ $p_d$-smooth, $k$ coprime with $N$, and $\gamma\in \bN\setminus\{0\}$, solutions of the Diophantine equation
\begin{equation}
 u =  v + kN^\gamma.\label{eq:conguse}
\end{equation}

By design, Schnorr's algorithm is only able to find solutions where $k$ is $p_d$-smooth and $\gamma = 1$ (Adleman's variant can yield, in principle, solutions with $\gamma>1$).  We look for pairs $(u,k)$ of $p_d$-smooth numbers satisfying the inequality 
\begin{equation}
 |u-kN|\leq p_d ,\label{eq:congineq}
\end{equation}
and we build solutions out of these pairs by setting $v = u-kN$: the inequality guarantees the $p_d$-smoothness of $v$. This search is lattice-based, and it involves lattice reduction and lattice enumeration algorithms.
%The elements of a well-chosen lattice close to a well-chosen point in space. It can be also seen as the search for many short vectors in a similar lattice.

Although in 1987 de Weger \cite{deWg87} had already applied lattice reduction to the effective resolution of Diophantine equations of the form (\ref{eq:conguse}), it was Schnorr who first applied it to factorization, in 1993 \cite{Sch93}. In 1995, Adleman \cite{Adl95} used Schnorr's approach to propose a reduction (not completely proved) from integer factorization to the search of a shortest nonzero vector in a lattice. Schnorr's algorithm was successfully implemented by Ritter and R\"{o}ssner in 1997 \cite{RR97}. 
%In 2009, the algorithm has been revisited by Schnorr himself \cite{Sch09}. 
%He adapted his original approach (in the $1$-norm) to the Euclidean norm, and proposed a new enumeration algorithm.

In this report, we improve a result of \cite{Sch93} by recycling a result of Micciancio \cite[Prop. 5.10]{MGBook}. This result may be useful (cf. remark \ref{remk:prove}) to show the existence of solutions to (\ref{eq:conguse}). In addition, we provide explicit computations of the volumes and the Gram-Schmidt Orthogonalizations of the involved lattices and lattice bases, respectively.

The road map is the following. First, in section \ref{sec:prime-number-lattice}, we introduce the lattice framework of Adleman, and we explain how can we solve the Diophantine equation (\ref{eq:conguse}) by searching short vectors in Adleman's lattice. Later in the same section, we explain the original approach of Schnorr, by particularizing Adleman's approach. Afterwards, in section \ref{sec:alg-computations} we give some properties of the Prime Number Lattices of Schnorr and Adleman. Finally, in section \ref{sec:conclusion}, we provide our conclusions and perspectives.

\section{Detecting solutions}
\label{sec:prime-number-lattice}
In this section we present the approaches of Adleman and Schnorr to solving (\ref{eq:conguse}) using lattices. We start by the approach of Adleman, which considers a search for short vectors. We show a sufficient condition to solving inequality (\ref{eq:congineq}). Then we present the approach of Schnorr, which considers a search for close vectors, and which can be seen as a particular case of Adleman's. We show a corresponding sufficient condition to solving (\ref{eq:congineq}).

\subsection{Coding a candidate solution}
\label{sec:coding-cand-solut}
Let $\nz\in\bZ^{d+1}$ be a vector with negative last coordinate. To this vector we associate a candidate solution to (\ref{eq:conguse}) in the following way 
\begin{equation}
u = \prod_{z_i>0, i\leq d}^d p_i^{z_i} ,\quad k= \prod_{z_i<0,i\leq d} p_i^{-z_i} \quad \text{and}\quad \gamma = |z_{d+1}|.\label{eq:defuv}
\end{equation}
Note that $u$ and $k$ are coprime. We would like to have candidate solutions providing an actual solution with high probability, that is, we want $v= u-kN^\gamma$ to be probably $p_d$-smooth. Now we will describe a way to find such candidate solutions.

\subsection{Making smoothness probable : the Prime Number Lattice of Adleman}
\label{sec:PNLA}
Define Adleman's $p$-norm Prime Number Lattice $\cA_p$ by the columns of the basis matrix 
$$  \nA_p =\left[
  \begin{array}{cccc}
\sqrt[p]{\ln p_1} & 0 & 0 & 0 \\
0 & \ddots & 0 & 0\\
0 & 0 & \sqrt[p]{\ln p_d} & 0 \\
C\ln p_1 & \cdots & C\ln p_d & C\ln N
\end{array}
\right],
$$
where $C>0$ is an arbitrary constant, which can depend on $N$.
%We will impose conditions on the $p$-norm of $\nA_p\nz$. Usually we will take $p=1$ or $2$, but we will perform the computations for the case of general $p$ whenever possible. 
The  vector $\nz\in \bZ^{d+1}$ satisfies
$$\nA_p\nz = \left[\begin{array}{c}
    z_1\sqrt[p]{\ln p_1} \\
    \vdots \\
     z_d\sqrt[p]{\ln p_d} \\
    C \left(\sum_{i=1}^dz_i\ln p_i + z_{d+1}\ln N\right)
  \end{array} \right]$$
and 
$$ ||\nA_p\nz||_p^p = \sum_{i=1}^d |z_i|^p\sqrt[p]{\ln p_i}^p + C^p\left|\sum_{i=1}^d z_i\ln p_i -|z_{d+1}|\ln N\right|^p, $$
and considering that this vector  codes a candidate solution, we have
$$ ||\nA_p\nz||_p^p = \sum_{i=1}^d |z_i|^p\ln p_i + C^p|\ln u -\ln (kN^{\gamma})|^p $$
and hence
$$ ||\nA_1\nz||_1 = \ln u + \ln k + C|\ln u -\ln (kN^{\gamma})|. $$
We have the following theorem in the case of the $1$-norm.
\begin{theorem}\label{th:svp-capture}
  Let $C>1$ and $\nz\in \bZ^{d+1}$, with $\gamma = |z_{d+1}|$ and $z_{d+1} < 0$. Then, whenever
  \begin{equation}
 \pnorm[1]{\nA_1\nz} \leq 2\ln C + 2\sigma\ln p_d - \gamma\cdot \ln N,\label{eq:1ball}
\end{equation}
we have
$$ |u-kN^\gamma| \leq p_d^\sigma. $$
\end{theorem}
\begin{proof}
  Just use lemma \ref{lemma:svp-suff} (in the appendix) with $\varepsilon = 2\ln C + 2\sigma\ln p_d - \gamma\cdot \ln N$.
\end{proof}
\begin{remk}\label{remk:negative}
  The requirement $z_{d+1} < 0$ is just needed to obtain a valid candidate solution. It does not reduce the space of solutions in any way, since a lattice is an additive group: for each vector of nonzero last coordinate, either itself or its opposite will have a strictly negative last coordinate.
\end{remk}
\begin{remk}
  When $\sigma =1$ and $\nz$ satisfies (\ref{eq:1ball}), we necessarily have a solution to the original equation
%\footnote{In fact, this is true at least for  $1\leq \sigma < \log_{p_d}(p_{d+1})$.}
  (\ref{eq:conguse}). In addition, when $\sigma>1$ is
  not too big, we can be quite optimistic about the $p_d$-smoothness
  of $v=u-kN^\gamma$, and hence on obtaining a solution too.
\end{remk}
\begin{remk}
  In order to factor $N$, one will typically search for (short) vectors $\nA_1\nz$ satisfying (\ref{eq:1ball}) for some $\sigma$ not too big, and then reconstruct from $\nz$ the candidate solution to (\ref{eq:conguse}), testing afterwards if it really constitutes a solution. In that case, the solution is stored, until we collect $d+2$ of them.
\end{remk}
\begin{remk}\label{remk:prove}
  Together with some extra knowledge on the properties of $\gamma$ for $\nz$ satisfying (\ref{eq:1ball}) (see remark \ref{remk:gamma}), theorem \ref{th:svp-capture} could be useful to prove the existence of solutions to inequality (\ref{eq:congineq}) and hence to equation (\ref{eq:conguse}), since we have explicit estimates on the length of a short nonzero vector of $\cA_1$, thanks to Minkowski's theorem for the $1$-norm. See Siegel \cite[Theorem 14]{Siegel89}.
\end{remk}
\begin{remk}
  Obtaining an analog of theorem \ref{th:svp-capture} for the Euclidean norm could be very useful, since this norm has better properties and it is the usual norm for lattice algorithms.
\end{remk}
%\begin{remk}
%  Usually one will take $\delta=\gamma$ on theorem \ref{th:svp-capture}, since otherwise a relatively big negative term on $C$ appears, potentially forcing (\ref{eq:1ball}) to be false for every reasonable choice of the parameters.
%\end{remk}
\subsection{A similar approach : the Prime Number Lattice of Schnorr}
\label{sec:PNLS}
The Prime Number Lattice of Schnorr $\cS_p$ is generated by the columns of the basis matrix 
\begin{equation}
 \nS_p = \left[\begin{array}{ccc}
\sqrt[p]{\ln p_1} & 0 & 0 \\
0 & \ddots & 0 \\
0 & 0 & \sqrt[p]{\ln p_d} \\
C\ln p_1 & \cdots & C\ln p_d
\end{array}
\right]. \label{eq:schbasis}
\end{equation}
The vector
\begin{equation}
 \nt = \left[
  \begin{array}{c}
    0 \\
    \vdots \\
    0 \\
    C \ln N
  \end{array}
\right]\label{eq:tvec}
\end{equation}
is the target vector of a close vector search in $\cS_p$, which replaces the short vector search of Adleman's approach. Schnorr's algorithm considers vectors $\nz\in\bZ^d$, to which it associates the candidate solution $(u,k,\gamma)$ to (\ref{eq:congineq}) with $u$ and $k$ defined exactly as in (\ref{eq:defuv}), and $\gamma =1$. We have
$$ \nS_p\nz - \nt = \left[
      \begin{array}{c}
        z_1 \sqrt[p]{\ln p_1} \\
        \vdots \\
        z_d \sqrt[p]{\ln p_d} \\
        C(\sum_{i=1}^d z_i \sqrt[p]{\ln p_i} \ln N)
      \end{array}
    \right], $$
and hence
$$ \pnorm{\nS_p\nz-\nt}^p = \sum_{i=1}^d |z_i|^p\ln p_i + C^p\left|\sum_{i=1}^d z_i \ln p_i - \ln N\right|^p. $$
The following theorem is the analog of theorem \ref{th:svp-capture}.
\begin{theorem}\label{th:cvp-capture}
  Let $C> 1$ and $\nz\in \bZ^{d}$. Hence, if
  \begin{equation}
 \pnorm[1]{\nS_1\nz - \nt} \leq 2\ln C + 2\sigma\ln p_d - \ln N, \label{eq:1ballcvp}
\end{equation}
then
$$ |u-kN| \leq p_d^\sigma. $$
\end{theorem}
\begin{proof}
  Just use lemma \ref{lemma:cvp-suff} with $\varepsilon = 2\ln C + 2\sigma\ln p_d - \ln N$.
\end{proof}
\begin{remk}\label{remk:gamma}
  In order to factor $N$, we should look for vectors of $\cS_1$ close to $\nt$. The main idea is that vectors satisfying (\ref{eq:1ballcvp}) for some $\sigma\geq 1$ not too big are more likely to provide candidate solutions which in turn will provide solutions to (\ref{eq:conguse}). Adleman's approach has the apparent advantage of having a larger search space, hence having a greater potential for finding solutions. In practice, this seems to be a disadvantage, since the solutions to (\ref{eq:conguse}) seem to be exactly those coming from Schnorr's approach too. Hence, in Adleman's approach one seems to search for many candidates that do not provide solutions. This could be related to the fact that the target vector $\nt$ does not belong to the real span of $\cS_1$: if the component of $\nt$ in the orthogonal complement of the span of $\cS_1$ is sufficiently big, any short vector in Adleman's lattice $\cA_1$ having nonzero last coordinate must have a last coordinate of absolute value equal to $1$, hence leading to the same solutions as Schnorr's lattice (see \cite[Chapter 4, Lemma 4.1]{MGBook} for a related discussion).
\end{remk}
\begin{remk}
  A great algorithmic advantage of the approach of Schnorr over that of Adleman is that the choice of the basis can be \emph{essentially independent of the number $N$}. For example, this will be the case if $C$ depends only on the size of $N$. This has the very important implication of allowing a precomputation on the basis (for example an HKZ reduction) valid for all numbers of some fixed size.
\end{remk}
\begin{remk}
  Proving the existence of solutions to (\ref{eq:1ballcvp}) seems harder in this case, since one needs a bound on the covering radius, which is less well understood than the first minimum.
\end{remk}
\begin{remk}
  Just as in the case of Adleman, obtaining an analog of theorem \ref{th:cvp-capture} for the Euclidean norm could be very useful. First attempts at finding this analog were stopped by involved computations.
\end{remk}
\section{Some properties of the Prime Number Lattices}
\label{sec:alg-computations}
We present some useful computations which extend those given by Micciancio and Goldwasser \cite[Chapter 5, section 2.3]{MGBook}.

\subsection{Volumes of the Prime Number Lattices}
Here we provide closed forms for the volumes of the $p$-norm Schnorr and Adleman lattices. This generalizes Proposition 5.9 of \cite{MGBook}, which considers only $p=2$.
\begin{remk}
  Recall that the volume of the lattice generated by the columns of a (not necessarily full rank) basis matrix $\mathbf{B}$ is
  $$ \vol(\cL(\mathbf{B})) = \sqrt{\left|\det(\mathbf{B}^T \cdot \mathbf{B}) \right|}, $$
which is exactly $\det(\mathbf{B})$ when $\mathbf{B}$ has full rank.
\end{remk}
\begin{theorem}
The volume of the $p$-norm Adleman lattice $\cA_p$, whose basis is
$$ \nA_p = \left[
  \begin{array}{cccc}
\sqrt[p]{\ln p_1} & 0 & 0 & 0 \\
0 & \ddots & 0 & 0\\
0 & 0 & \sqrt[p]{\ln p_d} & 0 \\
C\ln p_1 & \cdots & C\ln p_d & C\ln N
\end{array}
\right]$$
is given by 
$$ \vol(\cA_p) = C \ln N \cdot \prod_{i=1}^d \sqrt[p]{\ln p_i}. $$
Furthermore, the volume of the $p$-norm Schnorr lattice $\cS_p$, whose basis is 
$$ \nS_p = \left[
  \begin{array}{ccc}
\sqrt[p]{\ln p_1} & 0 & 0 \\
0 & \ddots & 0 \\
0 & 0 & \sqrt[p]{\ln p_d} \\
C\ln p_1 & \cdots & C\ln p_d
\end{array}
\right], $$
is given by
$$ \vol(\cS_p) =  \sqrt{1 + C^2\sum_{i=1}^d(\ln p_i)^{2-2/p}} \cdot\prod_{i=1}^d \sqrt[p]{\ln p_i}. $$
\end{theorem}
\begin{proof}
  The case of $\cA_p$ is trivial, as the basis matrix is lower triangular. Let us consider the case of $\cS_p$. It is easy to see that the volume of $\cS_p$ is a multilinear function of the columns of $\nS_p$. Hence, factoring out $\sqrt[p]{\ln p_i}$, $i\in\llbracket 1,d\rrbracket$  from the $i$-th column, we obtain
$$ \vol(\cS_p) = \sqrt{|\det(\nS_p^T \nS_p)|} = \sqrt{|\det(\hat\nS_p^T \hat\nS_p)|}\cdot \prod_{i=1}^d \sqrt[p]{\ln p_i} , $$
where $\hat\nS_p$ is of the form (\ref{eq:stdform}) (see lemma \ref{lemma:volgen} in the appendix) with 
$$ x_i = C\cdot (\ln p_i)^{1-1/p}.  $$
Lemma \ref{lemma:volgen} implies that 
$$   \sqrt{|\det(\hat\nS_p^T \hat\nS_p)|} =  \sqrt{1+\sum_{i=1}^d (C(\ln p_i)^{1-1/p})^2  } = \sqrt{1+C^2\sum_{i=1}^d (\ln p_i)^{2-2/p}  }, $$ 
which concludes the proof.
\end{proof}

\subsection{Explicit Gram-Schmidt Orthogonalization}
Here we give explicit expressions for the coefficients of the Gram-Schmidt Orthogonalization (GSO) of the set $\{\nb_1,\dots,\nb_d,\nt\}$ of columns of $\nS_p$, augmented by the target vector $\nt$ (or, equivalently, of the set of columns of $\nA_p$).
\begin{theorem}
  Consider the columns $\{\nb_i\}_{i=1}^d$ of Schnorr's Prime Number Lattice basis (\ref{eq:schbasis}), as well as the target vector $\nt$ defined in (\ref{eq:tvec}). The Gram-Schmidt Orthogonalization of $\{\nb_1,\dots,\nb_d,\nt\}$ involves the quantities 
$$ D_j = 1 + C^2\sum_{i=1}^j (\ln p_i)^{2-2/p}\quad 1\leq j \leq d $$
and is given by 
$$ (\nb_k^{\star})_i = \left\{
  \begin{array}{cc}
    -\frac{C^2 \ln p_k (\ln p_i)^{1-1/p}}{D_{k-1}}  & i<k \\
    (\ln p_k)^{1/p} & i=k \\
    0 & k<i<d+1 \\
    \frac{C\ln p_k}{D_{k-1}} & i=d+1
  \end{array}
\right. $$
and 
$$ (\nt^\star)_i = \left\{
  \begin{array}{cc}
    -\frac{C^2 (\ln N) (\ln p_i)^{1-1/p}}{D_{d}} & i<d+1 \\
    \frac{C(\ln N)}{D_{d}} & i=d+1
  \end{array}
\right.. $$
The corresponding Euclidean norms satisfy 
$$ \pnorm[2]{\nb_k^{\star}}^2 = (\ln p_k)^{2/p}\frac{D_k}{D_{k-1}}\qquad \pnorm[2]{\nt^\star}^2 = \frac{(C\ln N)^2}{D_d}. $$
Furthermore, the projection $\nt$ on the span of $\{\nb_1,\dots,\nb_d\}$, which is the effective target vector for the close vector search of Schnorr's algorithm, is given by
$$ (\nt - \nt^\star)_i = \left\{
  \begin{array}{cc}
    \frac{C^2 (\ln N) (\ln p_i)^{1-1/p}}{D_{d}} & i<d+1 \\
    \frac{C(\ln N)(D_d-1)}{D_{d}} & i=d+1
  \end{array}
\right.. $$
\end{theorem}
\begin{proof}
  The matrix having $\{\nb_1,\dots,\nb_d,\nt\}$ as columns is of the form (\ref{eq:stdform2}) (see lemma \ref{lemma:gsouse} in the appendix) with
$$ x_i = \sqrt[p]{\ln p_i}, \qquad y_i = C\cdot  \ln p_i \quad 1\leq i\leq d, $$
and 
$$ y_{d+1} = C\ln N . $$
Hence, using lemma \ref{lemma:gsouse}, we directly obtain the theorem.
\end{proof}
\begin{remk}
  The explicit value of $\pnorm[2]{\nt^\star}$ can be used to better understand the search for close vectors of Schnorr's algorithm. This is a consequence of the fact that $\nt$ does not belong to the span of $\{\nb_1,\dots,\nb_d\}$.
\end{remk}

\section{Conclusions and perspectives}
\label{sec:conclusion}
Using an idea of Micciancio, we presented partial but rigorous results advancing towards an \emph{effective} reduction from factorization to the search of short or close lattice vectors in the Prime Number Lattice of Adleman or Schnorr, respectively. These results, valid only for the $1$-norm, improve over those of Schnorr \cite[Theorem 2]{Sch93} by getting rid of asymptotically vanishing terms. Proving similar results for the Euclidean norm may be very useful, since it has much better properties than the $1$-norm and it is the natural choice for lattice algorithms\footnote{Although recently, in \cite[Theorem 2]{Sch10}, Schnorr restated \cite[Theorem 2]{Sch93} in the context of the Euclidean norm, this is essentially a generic restatement valid for every $p$-norm, $p\geq 1$, which still involves asymptotic terms.}.

Furthermore, we provided new properties of the Prime Number Lattices and their usual bases (in $p$-norm, $p\geq 1$), extending those of Micciancio \cite[Chapter 5, Section 2.3]{MGBook}. These properties could be useful to better understand the close vector search which takes place at the core of Schnorr's algorithm.

The next step of this work is to understand the distribution of lattice elements providing solutions to (\ref{eq:congineq}) or even (\ref{eq:conguse}), in order to choose on a well-grounded basis between enumeration algorithms (\cite{K83,FP83}) and random sampling algorithms (\cite{GPV08}, \cite{Kl00}), in the context of an effective implementation.

\subsection{Acknowledgements}
Thanks to Damien Stehl\'{e} for regular discussions and encouragement, as well as for many pointers to the relevant literature. Thanks to Guillaume Hanrot for useful discussions.

\bibliography{cado-lfs,biblioLLL,Others}
\bibliographystyle{acm}

\appendix

\section{Underlying lemmas}
\label{sec:long-proofs}

\subsection{Lemmas used in section \ref{sec:prime-number-lattice}}
The following two lemmas are elementary generalizations of a result of Micciancio \cite[Prop. 5.10]{MGBook}. 
\begin{lemma}\label{lemma:svp-suff}
  Let $C>1$ and let $\nz\in \bZ^{d+1}$ have negative last coordinate of module $\gamma = |z_{d+1}|\geq 1$, satisfying
$$ \pnorm[1]{\nA_1\nz} \leq \varepsilon . $$
Hence, we have
$$ |u-kN^\gamma| \leq \frac{N^{\frac{\gamma}{2}}}{C} \cdot\exp\left(\frac{\varepsilon}{2}\right). $$
\end{lemma}
\begin{proof}
The proof is essentially the same of Proposition 5.10 of \cite{MGBook}. We maximize $ |u-kN^\gamma| $ subject to the constraint
\begin{equation}
  ||\nA_1\nz||_1\leq \varepsilon.\label{eq:constraint}
\end{equation}
Since 
$$ ||\nA_1\nz||_1 = \ln u + \ln k + C|\ln u -\ln (kN^{\gamma})|, $$
the constraint (\ref{eq:constraint}) is symmetric in $u$ and $kN^\gamma$, and we can suppose without loss of generality that $ u\geq kN^\gamma $. Now, the constraint (\ref{eq:constraint}) can be rewritten as
$$ (C+1)\cdot \ln u - (C-1)\cdot \ln k \leq \varepsilon + C\gamma\cdot \ln N, $$
which implies
$$ u \leq k^{\frac{C-1}{C+1}}\cdot N^{\frac{C\gamma}{C+1}}\cdot \exp\left(\frac{\varepsilon}{C+1}\right). $$
Replacing this maximal value for $u$ in the objective function we get
\begin{equation}
  k^{\frac{C-1}{C+1}}\cdot N^{\frac{C\gamma}{C+1}}\cdot \exp\left(\frac{\varepsilon}{C+1}\right) - kN^\gamma.\label{eq:objf}
\end{equation}
Now, we optimize this last expression as a function of $k$. Differentiating (\ref{eq:objf}) with respect to $k$ we obtain
$$  \left(\frac{C-1}{C+1}\right)\cdot k^{-\frac{2}{C+1}} \cdot N^{\frac{C\gamma}{C+1}}\cdot \exp\left(\frac{\varepsilon}{C+1}\right) -N^\gamma $$
and hence the maximum is reached in the point
$$ k = \left(\frac{C-1}{C+1}\right)^{\frac{C+1}{2}} \cdot N^{-\frac{\gamma}{2}} \exp\left(\frac{\varepsilon}{2}\right). $$
%$$ k^{\frac{C-1}{C+1}}\cdot N^{\frac{C\gamma}{C+1}}\cdot \exp\left(\frac{\varepsilon}{C+1}\right) =  \left(\frac{C-1}{C+1}\right)^{\frac{C-1}{2}} \cdot N^{\frac{C}{2}(\gamma-\gamma)} \cdot N^{\frac{\gamma}{2}}\cdot\exp\left(\frac{\varepsilon}{2}\right) $$
The maximum of the original function is hence
$$ \left(\frac{C-1}{C+1}\right)^{\frac{C-1}{2}} \cdot N^{\frac{\gamma}{2}} \cdot\exp\left(\frac{\varepsilon}{2}\right)\cdot \left(\frac{2}{C+1}\right) $$
and as\footnote{When $x>1$, the function $f(x) = \left(\frac{x-1}{x+1}\right)^{\frac{x-1}{2}} \left(\frac{2x}{x+1}\right)  $ is monotonically decreasing, with $f(0^+)=1$.}
$$ \left(\frac{C-1}{C+1}\right)^{\frac{C-1}{2}}\cdot \left(\frac{2}{C+1}\right) \leq \frac{1}{C}  $$
for $C>1$, we conclude that
$$ |u-kN^\gamma|\leq \frac{N^{\frac{\gamma}{2}}}{C} \cdot\exp\left(\frac{\varepsilon}{2}\right),$$
as wished.
\end{proof}

 \begin{lemma}\label{lemma:cvp-suff}
  Let $C>1$ and let $\nz\in \bZ^{d}$ satisfying
$$ \pnorm[1]{\nS_1\nz - \nt} \leq \varepsilon . $$
Hence, 
$$ |u-kN| \leq \frac{\sqrt{N}}{C} \cdot\exp\left(\frac{\varepsilon}{2}\right). $$
\end{lemma}
\begin{proof}
  Just take $\gamma=1$ in the proof of lemma \ref{lemma:svp-suff}. 
\end{proof}

\subsection{Lemmas used in section \ref{sec:alg-computations}}
The following are general lemmas, maybe of independent interest. Lemma \ref{lemma:gsouse} could find an application in the context of knapsack lattice bases.
\begin{lemma}\label{lemma:volgen}
  The volume of the lattice $\cL$ generated by the columns of the matrix 
  \begin{equation}
 \nB = \left[
  \begin{array}{cccc}
    1 & 0 & 0 & 0 \\
    0 & 1 & 0 & 0 \\
    0 & 0 & \ddots & 0 \\
    0 & 0 & 0 & 1 \\
    x_1 & x_2 & \cdots & x_d
  \end{array}
\right]\label{eq:stdform}
\end{equation}
satisfies 
$$ \vol(\cL) = \sqrt{\det(\nB^T \nB)} = \sqrt{1+\sum_{i=1}^d x_i^2 }. $$
\end{lemma}
\begin{proof}
  We use Sylvester's determinant theorem (see for example \cite{MRMSylv}), which states that for every $\nA\in \bR^{m\times n}$ and $\nB\in \bR^{n\times m}$,
$$ \det(\nI_m + \nA\nB) = \det(\nI_n + \nB\nA), $$
where $\nI_k$ is the $k\times k$ identity matrix. Writing the matrix $\nB$ by blocks, and computing the associated Gram matrix, we obtain
$$ \nB = \left[
  \begin{array}{c}
    \nI_d \\
    \nx^T
  \end{array}
\right] \qquad \nB^T\nB = \nI_d + \nx\cdot \nx^T, $$
and hence, using Sylvester's theorem,
$$ \vol(\cL)^2 = \det(\nB^T\nB) = \det(\nI_d + \nx\cdot \nx^T) = \det(\nI_1 + \nx^T\cdot \nx) = 1 + \sum_{i=1}^d x_i^2, $$
as wished.
\end{proof}
\begin{lemma}\label{lemma:gsouse}
The Gram-Schmidt Orthogonalization of the columns $\{\nv_1,\dots, \nv_{d+1}\}$ of a nonsingular square matrix
\begin{equation}
 \left[
  \begin{array}{ccccc}
    x_1 & 0 & 0 & 0 & 0 \\
    0 & x_2 & 0 & 0 & 0 \\
    0 & 0 & \ddots & 0 & 0\\
    0 & 0 & 0 & x_d & 0\\
    y_1 & y_2 & \cdots & y_d & y_{d+1}
  \end{array}
\right]\label{eq:stdform2}
\end{equation}
can be specified in function of its entries and the quantities
$$ K_j = 1 + \sum_{i=1}^{j}\left(\frac{y_i}{x_i}\right)^2 \qquad 1\leq j\leq d,\quad K_0 =1, $$
by
\begin{equation}
 (\nv_k^\star)_i = \left\{
  \begin{array}{cc}
    -\left(\frac{ y_k}{K_{k-1}}\right)\cdot \left(\frac{y_i}{x_i}\right) & i<k \\
    x_k & i=k \\
    0 & k<i<d+1 \\
    \frac{y_k}{K_{k-1}} & i=d+1
  \end{array}
\right.\label{eq:gsv}
\end{equation}
for $k\leq d$, and by the same expression considering only the $i<k$ and $i=d+1$ cases, when $k=d+1$.
The Euclidean norms satisfy
\begin{equation}
 ||\nv_k^\star ||^2 = x_k^2 \frac{K_k}{K_{k-1}},\quad ||\nv_{d+1}^\star ||^2 = \frac{y_{d+1}^2}{K_d},\label{eq:enorm}
\end{equation}
and the Gram-Schmidt coefficients are 
\begin{equation}
 \mu_{k,j} = \frac{\nv_k\cdot \nv_j^\star}{\nv_j^\star\cdot \nv_j^\star} = \frac{y_k\cdot y_j}{x_j^2 K_j},  \quad 1\leq j<k\leq d+1. \label{eq:gsc}
\end{equation}
\end{lemma}
\begin{proof}
  The proof of (\ref{eq:gsv}) is carried out by induction. The result is clearly true for $k=1$. Suppose that it holds for $\nv_1^\star,\dots,\nv_{k-1}^\star$ for some $k\in \llbracket 2,d+1 \rrbracket$. Let us show that it still holds for $\nv_k^\star$. First, observe that for $1\leq j<k\leq d+1$, 
$$ \nv_k\cdot \nv_j^\star = (\nv_k)_{d+1}\cdot (\nv_j^\star)_{d+1} =  y_k\frac{y_j}{K_{j-1}} $$
and
\begin{eqnarray*}
  \pnorm[2]{\nv_j^\star}^2 = \nv_j^\star\cdot \nv_j^\star & = & \sum_{i=1}^{j-1}\left(\frac{y_i}{x_i}\right)^2\cdot \left(\frac{y_j}{K_{j-1}}\right)^2 + x_j^2 + \left(\frac{y_j}{K_{j-1}}\right)^2 \\
                                             & = & \left(\frac{y_j}{K_{j-1}}\right)^2\cdot\left( 1 + \sum_{i=1}^{j-1}\left(\frac{y_i}{x_i}\right)^2\right) + x_j^2 \\
                                             & = & \frac{y_j^2}{K_{j-1}} + x_j^2 \\
                                             & = & x_j^2\left( 1+ \frac{(y_j/x_j)^2}{K_{j-1}} \right) \\
                                             & = & x_j^2\left(\frac{K_{j-1} + (y_j/x_j)^2}{K_{j-1}} \right) \\                                             
                                             & = & x_j^2\frac{K_j}{K_{j-1}},
\end{eqnarray*}
which entails 
\begin{equation}
\mu_{k,j} = \frac{\nv_k\cdot \nv_j^\star}{\nv_j^\star\cdot \nv_j^\star} = \frac{y_k\cdot y_j}{x_j^2 K_j}.\label{eq:prgsc}
\end{equation}

Now, let $i\in \llbracket 1, k-1 \rrbracket $. By the definition of the Gram-Schmidt process, we have
\begin{eqnarray*}
  (\nv_k^\star)_i & = & (\nv_k)_i - \sum_{j=1}^{k-1} \mu_{k,j}\cdot(\nv_j^\star)_i \\
                        & = & 0 - \sum_{j=i}^{k-1} \mu_{k,j}\cdot(\nv_j^\star)_i \\
                        & = & - \mu_{k,i}\cdot(\nv_i^\star)_i - \sum_{j=i+1}^{k-1} \mu_{k,j}\cdot(\nv_j^\star)_i \\
                        & = & -\left(\frac{y_k y_i}{x_i^2 K_i}\right)\cdot x_i - \sum_{j=i+1}^{k-1} \left(\frac{y_k\cdot y_j}{x_j^2\cdot K_j}\right) \cdot \left(-\frac{y_i y_j}{x_i K_{j-1}}\right) \\
                        & = & -y_k \left(\frac{y_i}{x_i}\right) \left(\frac{1}{K_i} - \sum_{j=i+1}^{k-1} \left(\frac{y_j}{x_j}\right)^2 \frac{1}{K_{j-1}K_j} \right) \\
                        & = & -y_k \left(\frac{y_i}{x_i}\right) \left(\frac{1}{K_i} - \sum_{j=i+1}^{k-1} \left(\frac{1}{K_{j-1}} - \frac{1}{K_j}\right) \right) \\
                        & = & -\frac{y_k}{K_{k-1}} \left(\frac{y_i}{x_i}\right),
\end{eqnarray*}
as we wanted. Now, when $i=k\leq d$, 
\begin{eqnarray*}
   (\nv_k^\star)_k & = & (\nv_k)_k - \sum_{j=1}^{k-1} \mu_{k,j}\cdot(\nv_j^\star)_k \\
                          & = & x_k - \sum_{j=1}^{k-1} \mu_{k,j}\cdot 0 \\
                          & = & x_k,
\end{eqnarray*}
as we wanted. When $k<i\leq d$, we have
\begin{eqnarray*}
  (\nv_k^\star)_i & = & (\nv_k)_i - \sum_{j=1}^{k-1} \mu_{k,j}\cdot (\nv_j^\star)_i \\
                        & = & 0 - \sum_{j=1}^{k-1} \mu_{k,j} \cdot 0 \\
                        & = & 0
\end{eqnarray*}
as wished. Finally, when $i=d+1$ we obtain, for every $k\in \llbracket 2,d+1\rrbracket$, 
\begin{eqnarray*}
   (\nv_k^\star)_{d+1} & = & (\nv_k)_{d+1} - \sum_{j=1}^{k-1} \mu_{k,j}\cdot (\nv_j^\star)_{d+1} \\
                        & = & y_k - \sum_{j=1}^{k-1} \left(\frac{y_k y_j}{x_j^2 K_j}\right)\cdot \left(\frac{y_j}{K_{j-1}}\right) \\
                        & = & y_k \left( 1 - \sum_{j=1}^{k-1} \left(\frac{y_j}{x_j}\right)^2 \frac{1}{K_{j-1}K_j}  \right) \\
                        & = & y_k \left( 1 - \sum_{j=1}^{k-1} \left(\frac{1}{K_{j-1}} - \frac{1}{K_j}\right)  \right) \\
                        & = & y_k \left( 1 -\left(\frac{1}{K_0} - \frac{1}{K_{k-1}}\right)  \right) \\
                        & = & \frac{y_k}{K_{k-1}},
\end{eqnarray*}
since $K_0=1$. Hence, (\ref{eq:gsv}) is proved, both in the $1\leq k\leq d$ and the $k=d+1$ cases, as specified in the statement of the lemma. As a consequence of the computations preceding (\ref{eq:prgsc}), properties (\ref{eq:enorm}) and (\ref{eq:gsc}) are also proved, except for the Euclidean norm of $\nv_{d+1}^\star$, which is given by 
$$ \pnorm[2]{\nv_{d+1}^\star}^2 = \left( \frac{y_{d+1}}{K_d}\right)^2\cdot \left( 1 + \sum_{i=1}^d \left(\frac{y_i}{x_i}\right)^2 \right) = \frac{y_{d+1}^2}{K_d}. $$
The proof of the lemma is now complete.
\end{proof}

\end{document}